Comments on the Paper

# 'Earth's energy imbalance and implications'
# By J. Hansen, M. Sato, P. Kharecha, and K. von Schuckmann

**Gerhard Kramm**[1] and **Ralph Dlugi**[2]

[1]University of Alaska Fairbanks, Geophysical Institute

903 Koyukuk Drive, P.O. Box 757320, Fairbanks, AK 99775-7320, USA

[2]Arbeitsgruppe Atmosphärische Prozesse (AGAP),

Gernotstraße, D-80804 Munich, Germany

**Abstract:** In our comments we explicitly acknowledge the attempt of Hansen et al. to assess various uncertainties inherent in geophysical data being based on different measuring concepts and observation methods. However, with regard to the planetary energy budget, this paper offers some vulnerable points. We will focus our comments on these vulnerable points only. We will show that the energy imbalance of the entire Earth-atmosphere system is, indeed, based on these inherent uncertainties. We will demonstrate that the accuracy in the quantification of the global energy flux budget as claimed by Hansen et al. is, by far, not achievable in case of the entire Earth-atmosphere system. Using the value of the solar constant of $S_0 = 1361 \, \text{W m}^{-2}$ recently determined on the basis of total-solar-irradiance (TSI) observation by three different satellite projects (ACRIMSAT/ACRIM3 launched in 2000, SORCE/TIM launched in 2003, and PICARD/ PREMOS launched in 2010) we will document that the planetary energy imbalance of $F = 0.58 \pm 0.15 \, \text{W m}^{-2}$ calculated by Hansen et al. does not exist. Consequently, the implications related to this planetary energy imbalance have no basis.

## 1. Introduction

With great interest we read the paper of Hansen et al. (2011). We explicitly acknowledge the authors' attempt to assess various uncertainties inherent in geophysical data being based on different measuring concepts and observation methods. However, with regard to the planetary energy budget, this paper offers some vulnerable points. These points will be formulated and discussed in the following. In doing so, it is inevitable to demonstrate the interrelation between the so-called climate feedback equation and the climate sensitivity because the latter is related to the planetary energy imbalance assessed by Hansen et al. (2011) as an anthropogenic radiative forcing that is used in their simplified calculations of global temperature (see Section 2). In Section 3, we analyze the physical background of the global energy budget of the Earth-atmosphere system because diagnosing a planetary energy imbalance requires adequate energy flux budgets for both, the Earth's surface and the top of the atmosphere (TOA). The planetary energy imbalance, i.e., the net radiation at the TOA, as diagnosed by Hansen et al. (2011) is assessed in Section 4. Finally, we document that the accuracy in the data quality of the global energy flux budget as claimed by Hansen et al. (2011) is, by far, not achievable in case of the entire Earth-atmosphere system. Based on the value of the solar constant of $S_0 = 1361 \, \text{W m}^{-2}$ recently determined at the hand of three different satellite projects (ACRIMSAT/ACRIM3 launched in 2000, SORCE/TIM launched in 2003, and PICARD/ PREMOS launched in 2010, see, e.g., Kopp et al., 2012) it is shown that the planetary energy imbalance of $F = 0.58 \pm 0.15 \, \text{W m}^{-2}$ calculated by Hansen et al. (2011) does not exist. Consequently, the implications related to this planetary energy imbalance have no basis.

## 2. The interrelation between the climate feedback equation and the climate sensitivity

Hansen et al. (2011) employed climate forcings for simplified calculations of global temperature for demonstrating that a simple Green's function calculation, with negligible computation time, yields practically the same global temperature change as the complex climate model, provided that the global model's "climate response function" has been defined. Consequently, we assess the physical and mathematical framework on which their simplified calculations of global temperature are based. In doing so, we start with Eq. (1) of Hansen et al. (2011) given by

$$S = \frac{\Delta T_{eq}}{F} \quad . \tag{1}$$

Here, $S$ is the so-called climate sensitivity parameter, $\Delta T_{eq}$ is the change of the global surface temperature, $T_s$, from the equilibrium of the natural system to the system perturbed by the anthropogenic radiative forcing $F$. Equation (1) is based on the so-called climate feedback equation that has its origin in the global energy balance model of Schneider and Mass (1975) for the water layer of the thickness $\vartheta_w$ of an aqua planet. This global energy balance model reads

$$R \frac{dT_s}{dt} = (1 - \alpha_E) \frac{S_0}{4} - F_{IR,TOA}(T_s) \quad . \tag{2}$$

Here, $(1-\alpha_E) S_0/4 = \langle (1-\alpha(\Theta_0, \theta, \phi)) S \cos \Theta_0 \rangle_A$ is the solar radiation energetically relevant for the system Earth - atmosphere, $S_0$ is the solar constant, $\alpha(\Theta_0, \theta, \phi)$ is the position-dependent albedo in the solar range, $\alpha_E$ is the corresponding planetary albedo, $R = C_w \vartheta_w$ is the planetary inertia coefficient, where $C_w$ is the heat capacity of water, $T_s$ is the surface temperature, and $\cos \Theta_0 = \cos \theta \sin \delta + \sin \theta \cos \delta \cos h$ is the local zenith angle of the Sun's center, where $\delta$ is the solar declination angle, and $h$ is the hour angle from the local meridian. Furthermore, $\theta$ and $\phi$ are the zenith and azimuthal angles in the spherical coordinate frame, respectively. These angles characterize the given position. Moreover, the surface average of the globe, $\langle ... \rangle_A$, is defined by (e.g., Riley et al., 1998; Kramm et al., 2009)

$$\langle \Phi \rangle_A = \frac{r_E^2 \int_\Omega \Phi \, d\Omega}{r_E^2 \int_\Omega d\Omega} = \frac{1}{4\pi} \int_\Omega \Phi \, d\Omega \quad , \tag{3}$$

where $\Phi$ is an arbitrary quantity, $r_E$ is the radius of the globe, $\Omega = 4\pi$ is the solid angle of a sphere, and $d\Omega = \sin\theta \, d\theta \, d\phi$ is the differential solid angle.

The term on the left-hand side of Eq. (2) is not entirely correct. It must read $R \, dT_m/dt$, where $T_m = \langle T \rangle_V$ is the volume-averaged temperature for this layer (see Kramm and Dlugi, 2010). To replace $T_m$ by $T_s$ is generally invalid. However, if $dT_m/dt$ tends to zero like in case of a stationary state the different meaning of $T_m$ and $T_s$ becomes insignificant.

Following Schneider and Mass (1975), Eq. (2) should describe the change of the surface temperature of an aqua-planet with respect to time as a function of the energetically relevant solar radiation, $(1-\alpha_E)S/4$, and the outgoing infrared (IR) radiation counted for the top of the atmosphere (TOA), $F_{IR,TOA}$. The latter is given by

$$F_{IR,TOA} = F_{IR,a} + \tau_a \, F_{IR\uparrow} \quad , \tag{4}$$

where $F_{IR,a} = \langle F_{IR,a}(\theta, \phi) \rangle_A$ is the IR radiation emitted by the entire atmosphere in the direction of the space and reaching the TOA, $\tau_a \, F_{IR\uparrow} = \langle \tau_a(\theta, \phi) F_{IR\uparrow}(T_s(\theta, \phi)) \rangle_A$ is the IR radiation emitted by the Earth's surface, $\tau_a(\theta, \phi)$ is the position-dependent integral transmissivity of the atmosphere in the IR range, and $\tau_a$ is the corresponding planetary one. Note that the TOA may be interpreted as a height of the intervening atmospheric layer. Above this height neither solar radiation nor IR radiation is notably affected by atmospheric constituents.

Schneider and Mass (1975) expressed $F_{IR,TOA} = F_{IR,TOA}(T_s)$ as a function of the surface temperature using Budyko's (1969, 1977) empirical formula,

$$F_{IR,TOA} = a + b(T_s - T_r) - \{a_1 + b_1(T_s - T_r)\} \, n \quad . \tag{5}$$

Here, $a = 226.0 \text{ W m}^{-2}$, $b = 2.26 \text{ W m}^{-2} \text{ K}^{-1}$, $a_1 = 48.4 \text{ W m}^{-2}$, and $b_1 = 1.61 \text{ W m}^{-2} \text{ K}^{-1}$ are empirical coefficients, $T_r = 273.15 \text{ K}$ is the reference temperature, and $n$ is the normalized cloud cover. This means that this empirical formula (5) implies all radiative effects in the IR range, i.e., (a) the absorption and emission of IR radiation by the so-called greenhouse gases having either natural or anthropogenic origin, and (b) the IR radiation that is emitted by the Earth's surface and propagating through the atmosphere (it also includes the terrestrial radiation that is passing through the atmospheric window).

Inserting formula (5) into Eq. (2) and considering clear-sky conditions provides

$$R \frac{dT_s}{dt} = Q - \lambda T_s \quad , \tag{6}$$

with

$$Q = (1 - \alpha_E) \frac{S_0}{4} - a + b T_r \tag{7}$$

and $\lambda = S^{-1} = b$, i.e., the sensitivity parameter $S$ is the reciprocal of the so-called feedback parameter $\lambda$. Equation (6) is customarily called the climate feedback equation. If we interpret $T_s$ as a generalized coordinate and $dT_s/dt$ as the corresponding generalized velocity and assume that $Q$ is independent of time, we may transfer Eq. (6) into the phase space, where it is considered as a one-dimensional model because it has only one degree of freedom (Lange, 2007). As the feedback parameter, $\lambda$, is positive, the solution of Eq. (6) tends to an attractor given by $dT_s/dt = 0$, the condition of the fixed point (e.g., Kramm and Dlugi, 2010).

If we insert the anthropogenic radiative forcing $F$ into Eq. (7) relevant for clear-sky conditions, we will obtain

$$Q^{(p)} = (1 - \alpha_E) \frac{S}{4} - a + b T_r + F \quad , \tag{8}$$

where the superscript $(p)$ characterizes the perturbed case. Now, Eq. (6) can be solved using Eqs. (7) and (8) alternatively. Thus, we obtain two steady-state solutions: one for the natural system, and one for the system perturbed by F. The difference of these two solutions can be expressed by

$$\Delta T_{eq} = T_{eq}^{(p)} - T_{eq} = \frac{Q^{(p)}}{\lambda} - \frac{Q}{\lambda} = \frac{F}{\lambda} = S\,F \quad . \tag{9}$$

Rearranging this equation provides formula (1). Thus, a close interrelation between the climate feedback equation and the climate sensitivity exists (e.g., Dickinson, 1985; Berger and Tricot, 1986; National Research Council, 2005). The derivation of formula (1) already underlined the simplified character of the calculations addressed by Hansen et al. (2011). Nevertheless, it is indispensable to scrutinize the degree of this simplification and the inherent inaccuracy being performed in the next section.

## 3. Basic relations for the global energy budget of the Earth-atmosphere system

The global energy balance equation for the upper layer of an aqua-planet reads (see also Eq. (A17) of Kramm and Dlugi, 2010, and Figure 1)

$$R\frac{dT_m}{dt} = \left(1 - \alpha_E - A_a\right)\frac{S_0}{4} - H - E - H_w - \Delta F_{IR} \quad . \tag{10}$$

Here, $A_a\,S_0/4 = \left\langle A_a(\Theta_0, \theta, \phi)\,S\cos\Theta_0\right\rangle_A$ is the absorption of solar radiation by the entire atmosphere, where $A_a(\Theta_0, \theta, \phi)$ is the local absorption coefficient of the atmosphere in the solar range, $A_a$ is the corresponding planetary one, $H = \left\langle H(\theta, \phi)\right\rangle_A$ and $E = \left\langle E(\theta, \phi)\right\rangle_A$ are the fluxes of sensible and latent heat, respectively. Furthermore, the net radiation at the Earth's surface in the infrared range, $\Delta F_{IR}$, is given by

$$\Delta F_{IR} = \left\langle F_{IR\uparrow}\left(T_s(\theta,\phi)\right)\right\rangle_A - \left\langle \varepsilon_E(\theta,\phi) F_{IR\downarrow}(\theta,\phi)\right\rangle_A \quad, \tag{11}$$

where $F_{IR\uparrow}(T_s(\theta,\phi))$ is the emitted infrared radiation, $F_{IR\downarrow}(\theta,\phi)$ is the down-welling infrared radiation, and $\varepsilon_E(\theta,\phi) \leq 1$ is the emissivity at the Earth's surface. Moreover, $H_w = \left\langle H_w(\theta,\phi)\right\rangle_A$ represents the exchange of heat between the upper layer of the aqua-planet and deeper water layers and/or the ocean floor, respectively. Again, the zenith and azimuthal angles, $\theta$ and $\phi$, characterize the given position. The heat exchange, $H_w$, is usually ignored so that Eq. (10) becomes

$$R \frac{dT_m}{dt} = \left(1 - \alpha_E - A_a\right)\frac{S_0}{4} - H - E - \Delta F_{IR} \quad. \tag{12}$$

However, in case of the water layer under study $H_w$ might be contributing to an energy imbalance.

A similar equation can be derived for the atmosphere (see also Eq. (A18) of Kramm and Dlugi, 2010):

$$\vartheta_a \frac{d}{dt}\left\langle C_a T_a\right\rangle_V = A_a \frac{S_0}{4} - F_{IR,TOA} + H + E + \Delta F_{IR\uparrow} \quad, \tag{13}$$

where $\vartheta_a$ is the thickness of intervening atmospheric layer (i.e., between the Earth's surface and the TOA), $T_a$ is the atmospheric temperature, and $C_a$ is the heat capacity. Similar to Eq. (10) we might express the term of the left-hand side of this equation by

$$\vartheta_a \frac{d}{dt}\left\langle C_a T_a\right\rangle_V = C_a \vartheta_a \frac{d}{dt}\left\langle T_a\right\rangle_V = R_a \frac{d}{dt}\left\langle T_a\right\rangle_V \quad, \tag{14}$$

but this is a crude simplification. Such simplifications may play a role in lecture rooms to describe some effects qualitatively, but they have to be avoided in real scientific studies because they are, by far, not fulfilled.

Comparing Eq. (12) with Eq. (2) yields:

$$F_{IR,TOA} = A_a \frac{S_0}{4} + H + E + \Delta F_{IR} \quad . \tag{15}$$

This relation is only valid if the following condition is fulfilled:

$$\vartheta_a \frac{d}{dt}\langle C_a T_a \rangle_V = 0 = A_a \frac{S_0}{4} - F_{IR,TOA} + H + E + \Delta F_{IR} \quad . \tag{16}$$

This means that the atmosphere must always be in a stationary state. It is unlikely that this condition is generally fulfilled. Nevertheless, if the condition (16) is inserted into Eq. (10), one will obtain (see also Siegenthaler and Oeschger, 1984; Berger and Tricot, 1986):

$$R \frac{dT_m}{dt} = (1 - \alpha_E) \frac{S_0}{4} - F_{IR,TOA} - H_w \quad . \tag{17}$$

Obviously, ignoring $H_w$ as mentioned before leads to an expression similar to Eq. (2).

If the planetary radiation balance at the TOA is fulfilled as suggested by Trenberth et al. (2009) and many others (see Table 1), i.e.,

$$(1 - \alpha_E) \frac{S_0}{4} - F_{IR,TOA} = 0 \quad , \tag{18}$$

then Eq. (12) will read:

$$0 = (1 - \alpha_E - A_a) \frac{S_0}{4} - H - E - \Delta F_{IR} \quad . \tag{19}$$

In case of long-term global averages this energy flux budget for the Earth's surface is fulfilled, but a large scatter exists (see Table 1). Here, it is indispensably to cite Fortak (1971). In his book entitled "Meteorologie" he stated:

»The "cycle" of the long-wave radiation between that Earth's surface and the atmosphere does not contribute to the heating of the system. The effective outgoing emission of infrared radiation of 64 % only serves to maintain the radiative equilibrium at the top of the atmosphere.«

Note the value of 64 % has to be updated to 70 % or so for both the energetically relevant solar radiation and the outgoing IR radiation.

In case of Eq. (17) the planetary energy balance (18) leads to

$$R \frac{dT_m}{dt} = -H_w \quad . \tag{20}$$

Under such a condition the temporal change of the mean temperature of the upper layer of an aqua planet is related to the exchange of heat between this layer and deeper water layers and/or the ocean floor, respectively.

### 4. On the radiative imbalance at the TOA

First of all, it is indispensable to explain the notion "planetary energy imbalance" used by Hansen et al. (2011). This quantity was not explicitly defined by these authors, but they stated:

»The planetary energy imbalance (Hansen et al., 1997, 2005) is the fundamental relevant quantity, because it is a direct consequence of the net climate forcing.«

In the paper of Hansen et al. (2005) the planetary energy imbalance is related to the net radiation at the TOA. Hansen et al. (2005) stated:

»We infer from the consistency of observed and modeled planetary energy gains that the forcing still driving climate change, i.e., the forcing not yet responded to, averaged ~ 0.75 W/m² in the past decade and was ~ 0.85 ± 0.15 W/m² in 2003 (Fig. 1C).«

Without any doubt, Figure 1C of Hansen et al. (2005), here repeated as Figure 2 for the purpose of convenience, shows the net radiation at the TOA derived from climate simulations. It is similar to the diagram of the right-hand side of Figure 1 of Hansen et al. (2011). It should be noticed that Hansen et al. (2007) used for their climate simulations for the period 1880 – 2003 with the GISS modelE TSI values of Lean (2000) and Lean et al. (2002) here illustrated in Figures 3c and 3d, respectively. This means that the numerical simulations performed by Hansen et al. (2002, 2007) are based on a solar constant similar to that shown in Figure 4. Furthermore, the net radiation at the TOA presented by Hansen et al. (2007) is nearly identical with that illustrated in the Figure 1C of Hansen et al. (2005), here repeated as Figure 2. Thus, the planetary energy imbalance of $F = 0.85 \pm 0.15 \, W \, m^{-2}$ in 2003 as diagnosed by Hansen et al. (2005) is based on the same obsolete solar constant.

If the outgoing infrared radiation is reduced by $F$ due to the anthropogenic effect, Eq. (15) may be written as

$$F_{IR,TOA}^{(p)} = F_{IR,TOA} - F = A_a \frac{S_0}{4} + H + E + \Delta F_{IR} - F \quad , \tag{21}$$

where, again, the superscript $(p)$ characterizes the perturbed system. This would mean, for instance, that the net radiation at the Earth's surface is reduced by $F$, suggesting, in principle, an increasing down-welling infrared radiation,

$$\Delta F_{IR} - F = \langle F_{IR\uparrow}(T_s) \rangle_A - \{ \langle \varepsilon_E \, F_{IR\downarrow} \rangle_A + F \} \quad . \tag{22}$$

However, this does not automatically mean that the surface temperature must increase, as suggested by Eq. (1), to re-establish a planetary radiation balance at the TOA, characterized by Eq. (18), as already pointed out by Ramanathan et al. (1987) and now argued by Hansen et al. (2011) in a similar manner:

> »The basic physics underlying this global warming, the greenhouse effect, is simple. An increase of gases such as $CO_2$ makes the atmosphere more opaque at infrared wavelengths. This added opacity causes the planet's heat radiation to space to arise from higher, colder levels in the atmosphere, thus reducing emission of heat energy to space. The temporary imbalance between the energy absorbed from the Sun and heat emission to space, causes the planet to warm until planetary energy balance is restored.«

This argument is physically incorrect because none of these flux terms on the right-hand side of Eq. (12), i.e., $(1 - \alpha_E - A_a) S_0/4$, $H$, $E$, and $\Delta F_{IR}$, depends on the globally averaged surface temperature. Furthermore, there is no constant ratio between $H + E$ on the one hand and $\Delta F_{IR}$ on the other hand (see Table 1). A reduction of $\Delta F_{IR}$ by $F$ may easily be compensated by $H$ and/or $E$ to fulfill the requirement of the energy flux budget (19). The same is true in case of any other of these flux terms. Note that even the Bowen ratio $B = H/E$ is not constant (see Table 1). The uncertainty inherent in the determination of these fluxes is so large that $F$ may be assessed as noise in the energy flux budget for the Earth's surface (e.g., Kramm and Dlugi, 2009).

In the capture of their Figure 17 (repeated here for the purpose of convenience as Figure 4) that illustrates the composite of Fröhlich and Lean (1998) and the Physilkalisch-Meteorologisches Observatorium Davos (PMOD), World Radiation Center (WRC) for the last three decades, but excludes recent satellite observations, Hansen et al. (2011) stated:

»Recent estimates of mean solar irradiance (Kopp and Lean, 2011) are smaller, $1360.8 \pm 0.5$ W m$^{-2}$, but the uncertainty of the absolute value has no significant effect on the solar forcing, which depends on the temporal change of irradiance.«

Since Hansen et al. (2007) used for their climate simulations the reconstructed total solar irradiance (TSI) of Lean (2000) and Lean et al. (2002) that shows a variation smaller than 3 W m$^{-2}$ during the respective period of about 150 years (see Figure 3c and 3d), this statement seems to be awkward because this variation is notably smaller than the difference between the obsolete solar constant of about $S_0 = 1366$ W m$^{-2}$ illustrated in the Figure 17 of Hansen et al. (2011) and the current value of about $S_0 = 1361$ W m$^{-2}$. Note that this statement was not included in earlier versions of the manuscript of Hansen et al. from May 5 (see their Figure 21) and September 2, 2011 (see their Figure 17) that can be found in the arXiv of the Cornell University (http://arxiv.org/abs/1105.1140). Obviously, it is the authors' response to the paper of Kopp and Lean (2011).

In their Figure 17, Hansen et al. (2011) also claimed that before 1978 only proxy data for the solar constant were available, a repetition of a statement of Hansen et al. (2002), and that the SORCE data are only available since 2010. These claims are incorrect. First of all, SORCE/TIM was launched in 2003, and Butler et al. (2008) already reported that SORCE/TIM provided a solar constant of about $S_0 = 1361$ W m$^{-2}$. Second, based on the first direct observations of the solar constant carried out at an altitude of 82 km using an X-15 rocket aircraft, Laue and Drummond (1968) found: $S_0 = 1361$ W m$^{-2}$. By considering the 1969-1970 Nimbus3 observations, Raschke et al. (1973) estimated the solar constant by $S_0 \cong 1.952$ cal m$^{-2}$ min$^{-1} \approx 1361$ W m$^{-2}$. The same value was already suggested by Henry G. Houghton in a letter to the Journal of Meteorology in March 1951. His suggestion also included an ultraviolet correction that was based on the ultraviolet solar spectrum obtained from a V-2 rocket above the ozone layer (Hulbert, 1947).

As illustrated in Figure 5, during the past three decades satellite-derived solar constants "decreased" from $S_0 \cong 1371 \text{ W m}^{-2}$ (NIMBUS7/ERB, House et al. 1986) for the 5-year period November 1978-October 1983 and $S_0 \cong 1365 \text{ W m}^{-2}$ (ERBE/ERBS, Barkstrom et al., 1990) to the recent value of $S_0 \cong 1361 \text{ W m}^{-2}$ (ACRIMSAT/ACRIM3 launched in 2000, SORCE/TIM launched in 2003, and PICARD/ PREMOS launched in 2010; see Kopp and Lean, 2011, Kopp et al., 2012). However, as illustrated by Lean (2010) in her Figure 5 (repeated here for the purpose of convenience as Figure 6), some proposed TSI proxy reconstructions that slightly differ from each other only reflect small variations for the period of the past six decades. Thus, we may conclude that the solar constant was about $1361 \text{ W m}^{-2}$ during this past period.

The argument of Hansen et al. (2011) cited before that the correct value of the solar constant is of minor importance has to be discussed with respect to the recently measured and better quality-controlled values of the TSI.

If $F_{IR,TOA}$ is reduced by $F$ (see Eq. (8)) generating a planetary radiative imbalance at the TOA, Eq. (18) has to be written as

$$(1 - \alpha_E) \frac{S_0}{4} - F_{IR,TOA}^{(p)} = F \quad . \tag{23}$$

In contrast to the statement of Hansen et al. (2011), this equation documents that the correct absolute value of the solar constant is indispensable to determine the planetary energy imbalance, i.e., the net radiation at the TOA. It also documents that the planetary albedo plays a notable role if the planetary energy imbalance has to be determined. Since the planetary albedo is varying with time (see Figure 7), the solar radiation absorbed by the system Earth-atmosphere is also varying with time even in case of a nearly constant TSI value at the TOA. Consequently, also the argument of Duffy et al. (2009),

»Because incoming solar energy – that is, TSI – has not changed, the imbalance must result from increased absorption of outgoing energy, such as by increased greenhouse gases«, is rather inadequate. Note that these authors completely ignored the observational evidence provided by SORCE/TIM since 2003 (see Figure 8).

Obviously, the planetary energy imbalance of $F = 0.58 \pm 0.15 \, W \, m^{-2}$ for the period 2005-2010 deduced by Hansen et al. (2011, see subsection 14.4) is notably smaller than that of the quantity

$$0.25 \left(1 - \alpha_E\right) \left| S_{0,old} - S_{0,new} \right| \approx 0.88 \, W \, m^{-2} \qquad (24)$$

if a planetary albedo of $\alpha_E = 0.3$ is assumed as suggested by Figure 7. Here, $S_{0,old} = 1366 \, W \, m^{-2}$ and $S_{0,new} = 1361 \, W \, m^{-2}$. Note that the variability illustrated in Figure 17 of Hansen et al. (2011) is only related to $S_{0,old}$ (see Figure 4). This variability does not characterize the systematic difference $S_{0,old} - S_{0,new}$. This difference has to be assessed as a "procedural" error. Consequently, one may argue that in case of the youngest result for the solar constant the outgoing IR radiation would slightly be larger than the energetically relevant solar radiation. Thus, F would become negative or vanish, i.e., - according to Eq. (9) - $\Delta T_{eq} \approx 0$. Based on this result, the planetary energy imbalance claimed by Hansen et al. (2011) is not justifiable.

Finally, it is worthy to take a look on the $240 \, W \, m^{-2}$, according to Hansen et al. (2011, see subsection 13.2) the solar energy averaged over the planet's surface. First of all, if a planetary albedo of $\alpha_E = 0.3$ and a solar constant of $S_{0,new} = 1361 \, W \, m^{-2}$ are assumed, the energetically relevant solar radiation of $\left(1 - \alpha_E\right) S_0 / 4 \cong 238 \, W \, m^{-2}$ affects the entire earth-atmosphere system. Second, according to Trenberth et al. (2009) and many others, the solar radiation reaching the Earth's surface, $\left(1 - \alpha_E - A_a\right) S_0 / 4 \cong 160 \, W \, m^{-2}$, is much smaller because a portion of the solar irradiance at the TOA is already absorbed by the atmosphere ( $A_a \, S_0 / 4 \cong 78 \, W \, m^{-2}$ if $A_a = 0.23$ is assumed; see Table 1). If, again, a value of $\alpha_E = 0.3$ is used, the solar constant that corresponds to $S_0 = 4 \cdot 240 \text{ W m}^{-2}/(1-\alpha_E)$ must be $S_0 = 1371 \text{ W m}^{-2}$. This is the TSI which was delivered by early satellite observations starting with Nimbus7 in 1978 (see Figure 5). Additionally, a slight variation of the planetary albedo by $\Delta\alpha_E = \pm 0.01$ (see Figure 7) leads to a change in the energetically relevant solar radiation by $\pm 3.4 \text{ W m}^{-2}$ if $S_0 = 1361 \text{ W m}^{-2}$ is assumed.

## 5. Final remarks and conclusions

Kopp and Lean (2011) already stated:

> »Instrument inaccuracies are a significant source of uncertainty in determining Earth's energy balance from space-based measurements of incoming and reflected solar radiation and outgoing terrestrial thermal radiation. A nonzero average global net radiation at the top of the atmosphere is indicative of Earth's thermal disequilibrium imposed by climate forcing. But whereas the current planetary imbalance is nominally 0.85 W m$^{-2}$ [Hansen et al., 2005], estimates of this quantity from space-based measurements range from 3 to 7 W m$^{-2}$. SORCE/TIM's lower TSI value reduces this discrepancy by 1 W m$^{-2}$ [Loeb et al., 2009]. We note that the difference between the new lower TIM value with earlier TSI measurements corresponds to an equivalent climate forcing of −0.8 W m$^{-2}$, which is comparable to the current energy imbalance.«

This means that Hansen et al. (2005, 2011) have already overdone their estimates of the planetary energy imbalance. Based on available evidence, we, therefore, conclude that the accuracy in the quantification of the global energy flux budget as claimed by Hansen et al. (2011) is, by far, not achievable in case of the entire Earth-atmosphere system. Based on Eq. (28)

we may conclude that a planetary energy imbalance of $F = 0.58 \pm 0.15 \text{W m}^{-2}$ claimed by Hansen et al. (2011) for the period 2005-2010 is not justifiable. The same is true in case of the planetary energy imbalance of $F = 0.85 \pm 0.15 \text{W m}^{-2}$ claimed by Hansen et al. (2005). Consequently, the implications related to these planetary energy imbalances have to basis.

Siegenthaler, U., and Oeschger, H.: Transient temperature changes due to increasing $CO_2$ using simple models, Ann. Glaciology, 5, 153-159, 1984.

Tapping, K.F., Boteler, D., Charbonneau, P., Crouch, A., Manson, A., and Paquette, H: Solar magnetic activity and total irradiance since the maunder minimum, Solar Physics, 246, 309-326, 2007.

Trenberth, K.E., Fasullo, J.T., and Kiehl, J.: Earth's global energy budget, Bull. Amer. Met. Soc., 311-323, 2009.

United States Committee for the Global Atmospheric Research Program, Understanding climatic change: a program for action, National Academy of Sciences, Washington, 1975.

Vardavas, I.M., and Taylor F.W.: Radiation and climate, Oxford University Press, Oxford, U.K., 2007.

Wang, Y.-M., Lean, J.L., and Sheeley, J., N.R.: Modeling the Sun's magnetic field and irradiance since 1713, The Astrophysical Journal, 625, 522–538, 2005.
21

**Table 1:** Summary of the Earth's energy budget estimates (with respect to Kiehl and Trenberth, 1997). The sources [1], [2], and [15] are inserted, and source [9] is updated (adopted from Kramm and Dlugi, 2011).

| Earth's surface | | | | Atmosphere | TOA | Source |
|---|---|---|---|---|---|---|
| $(1-\alpha_E - A_a)S_0/4$ | $\Delta R_{L\uparrow}$ | H | E | $A_a$ | $\alpha_E$ | |
| 145 | 47 | 20 | 78 | 0.22 | 0.35 | [1] |
| 164 | 70 | 17 | 77 | 0.17 | 0.36 | [2] |
| 174 | 72 | 24 | 79 | 0.19 | 0.30 | [3] |
| 157 | 52 | 17 | 88 | 0.24 | 0.30 | [4] |
| 174 | 68 | 27 | 79 | 0.19 | 0.30 | [5] |
| 171 | 72 | 17 | 82 | 0.20 | 0.30 | [6] |
| 169 | 63 | 16 | 90 | 0.20 | 0.31 | [7] |
| 154 | 55 | 17 | 82 | 0.25 | 0.30 | [8] |
| 161 | 66 | 26 | 69 | 0.23 | 0.30 | [9] |
| 171 | 68 | 21 | 82 | 0.20 | 0.30 | [10] |
| 157 | 51 | 24 | 82 | 0.23 | 0.31 | [11] |
| 171 | 68 | 24 | 79 | 0.20 | 0.30 | [12] |
| 168 | 66 | 24 | 78 | 0.20 | 0.31 | [13] |
| 165 | 46 | - | - | 0.19 | 0.33 | [14] |
| 161 | 63 | 17 | 80 | 0.23 | 0.30 | [15] |

[1] Haltiner and Martin (1957), [2] Fortak (1971), [3] United States Committee for the Global Atmospheric Research Program (1975), [4] Budyko (1982), [5] Paltridge and Platt (1976), [6] Hartmann (1994), [7] Ramanathan (1987), [8] Schneider (1987), [9] Liou (2002), [10] Peixoto and Oort (1992), [11] MacCracken (1985), [12] Henderson-Sellers and Robinson (1986), [13] Kiehl and Trenberth (1997), [14] Rossow and Zhang (1995) ,and [15] Trenberth et al. (2009).

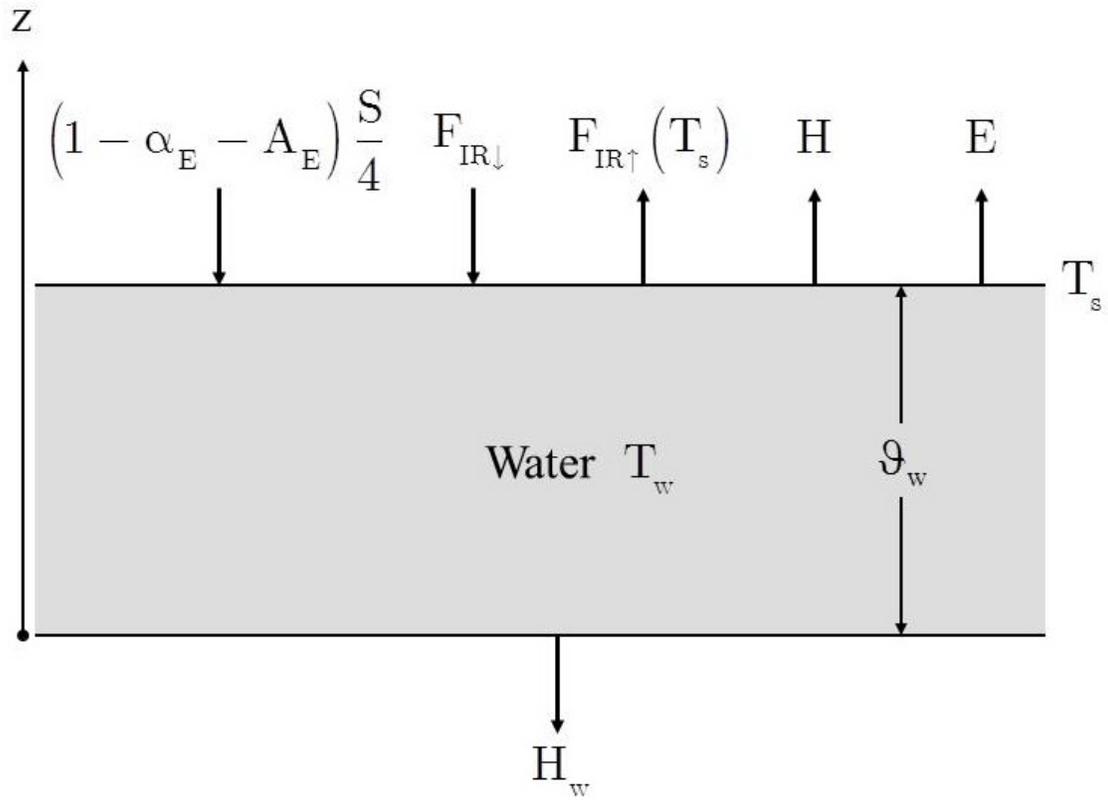

**Figure 1.** Sketch of the global energy flux budget for the upper layer of an aqua-planet (adopted from Kramm and Dlugi, 2010).

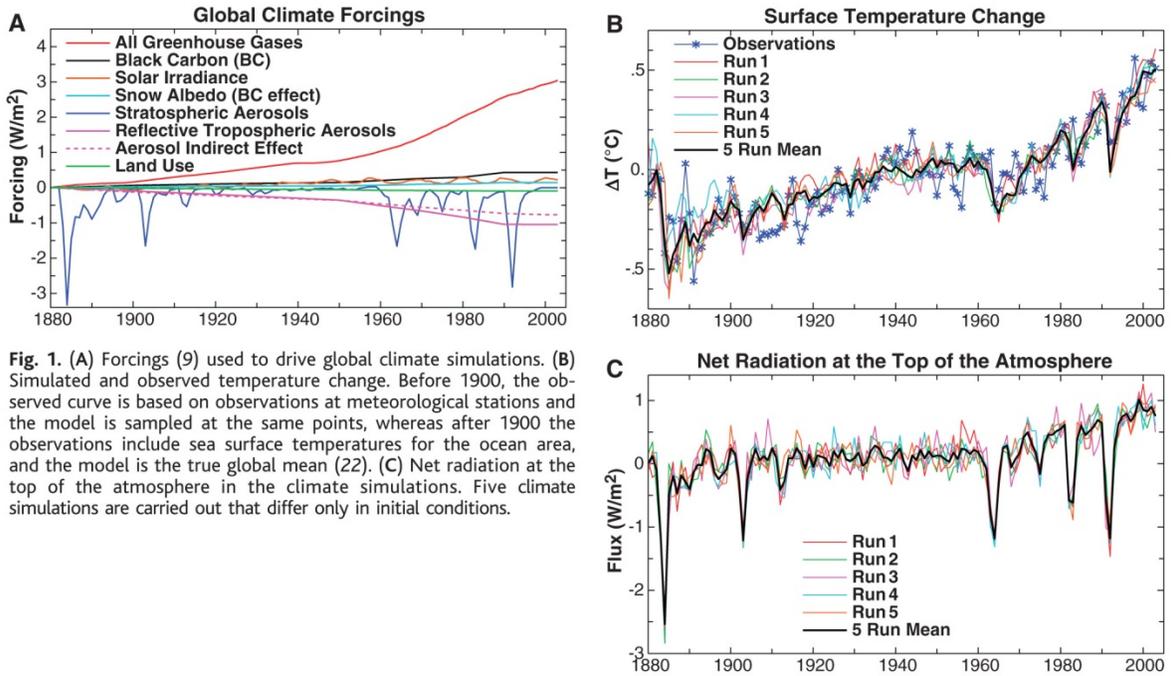

**Figure 2.** Shown is the Figure 1 of Hansen et al. (2005).

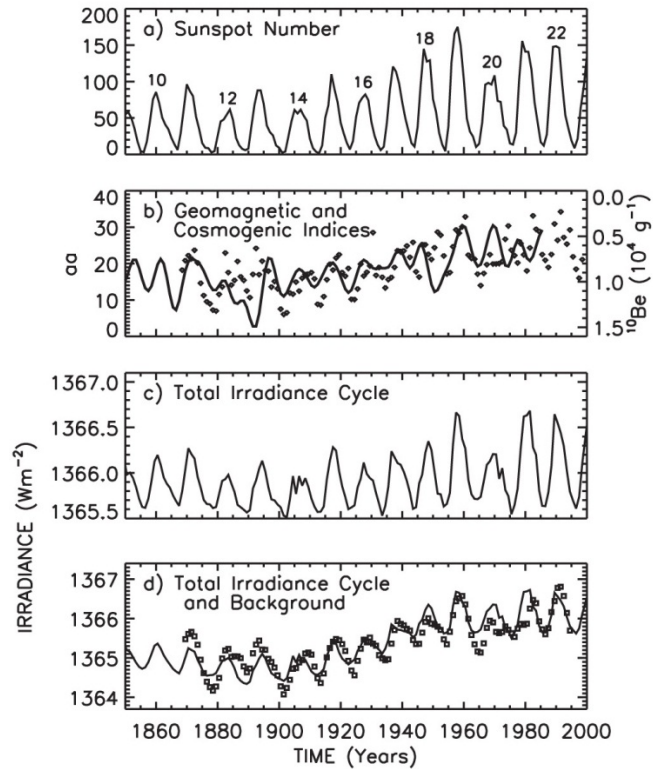

**Figure 3**. Shown in (a) are variations in the sunspot number compared in (b) with variations in terrestrial proxies of solar activity. The solid line is $^{10}$Be (Beer et al., 1988) and the symbols are the aa index. In (c) is reconstruction of total solar irradiance arising from 11-year activity cycles, whereas in (d) the irradiance reconstructions assume an additional varying background component. The solid line is from Lean (2000) and the symbols are from Lockwood and Stamper (1999). The sunspot numbers and aa indices were obtained from NOAA's National Geophysical Data Center (adopted from Lean et al., 2002).

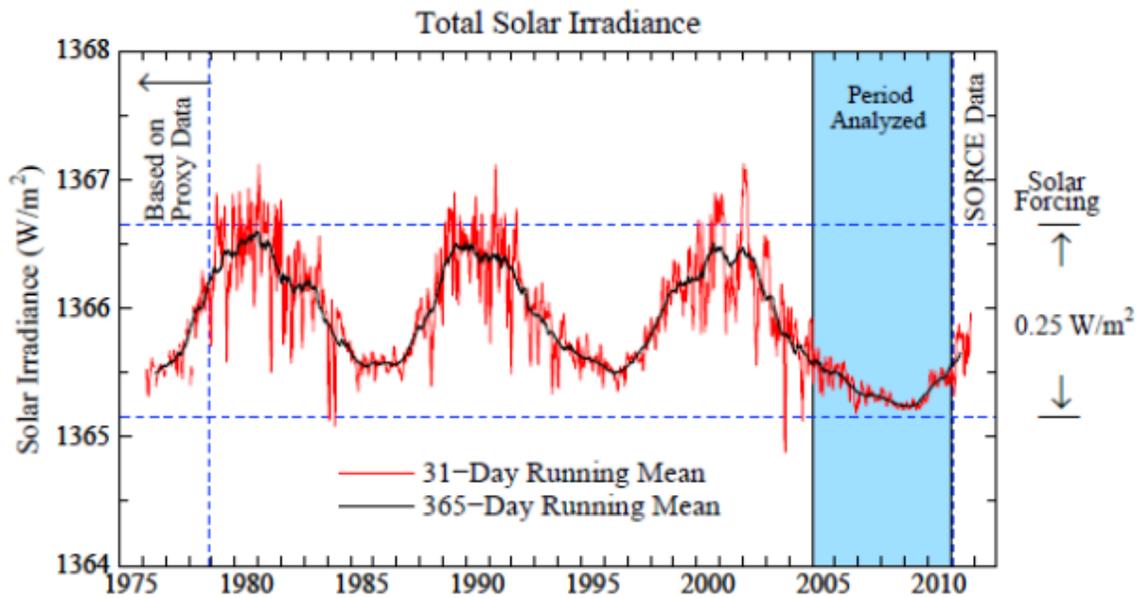

**Figure 4.** Solar irradiance from composite of several satellite-measured time series. Data through 2 February 2011 is from Fröhlich and Lean (1998 and Physikalisch Meteorologisches Observatorium Davos, World Radiation Center). Update in 2011 (through 24 August) is from University of Colorado Solar Radiation & Climate Experiment normalized to match means over the final 12 months of the Fröhlich and Lean data. Recent estimates of mean solar irradiance (Kopp and Lean, 2011) are smaller, $1360.8 \pm 0.5$ W m$^{-2}$, but the uncertainty of the absolute value has no significant effect on the solar forcing, which depends on the temporal change of irradiance (adopted from Hansen et al., 2011).

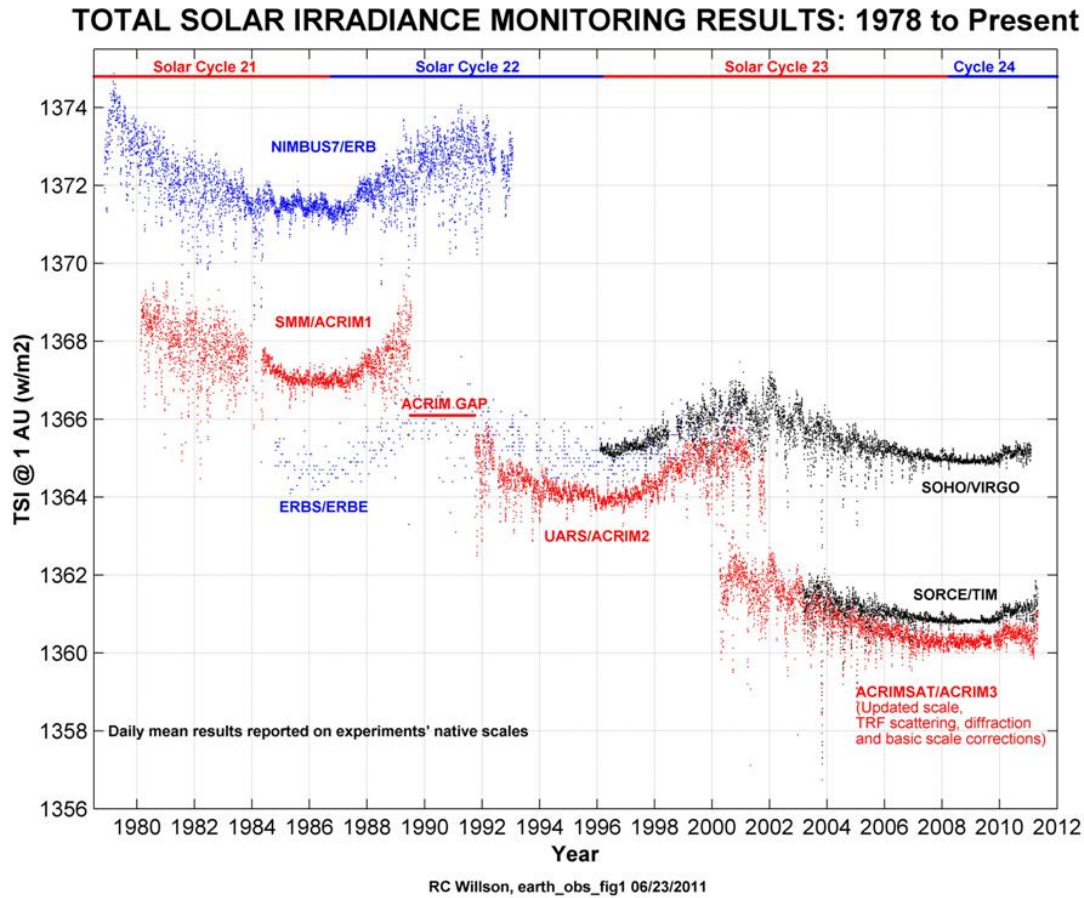

**Figure 5:** Satellite observations of total solar irradiance. It comprises of the observations of seven independent experiments: (a) Nimbus7/Earth Radiation Budget experiment (1978 - 1993), (b) Solar Maximum Mission/Active Cavity Radiometer Irradiance Monitor 1 (1980 - 1989), (c) Earth Radiation Budget Satellite/Earth Radiation Budget Experiment (1984 - 1999), (d) Upper Atmosphere Research Satellite/Active cavity Radiometer Irradiance Monitor 2 (1991 - 2001), (e) Solar and Heliospheric Observer/Variability of solar Irradiance and Gravity Oscillations (launched in 1996), (f) ACRIM Satellite/Active cavity Radiometer Irradiance Monitor 3 (launched in 2000), and (g) Solar Radiation and Climate Experiment/Total Irradiance Monitor (launched in 2003). The figure is based on Dr. Richard C. Willson's earth_obs_fig1, updated on June 23, 2011 (see http://www.acrim.com/).

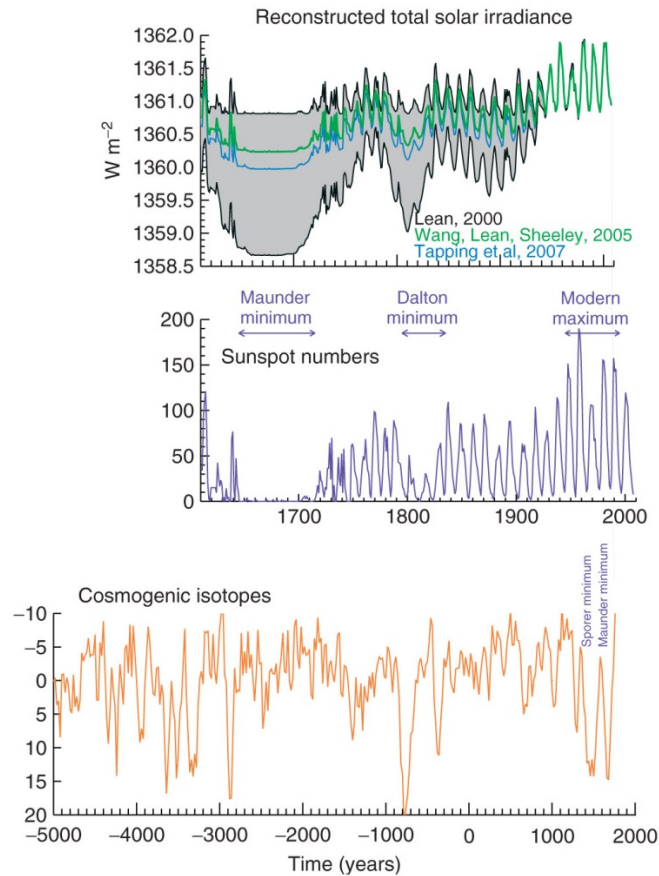

**Figure 6**. Reconstructions of total solar irradiance with different assumptions about the strength of the background component that underlies the activity cycle are compared since the Maunder Minimum (adopted from Lean, 2010). Shown are the corrected ones of Lean (2000), Wang et al. (2005), and Tapping et al. (2007).

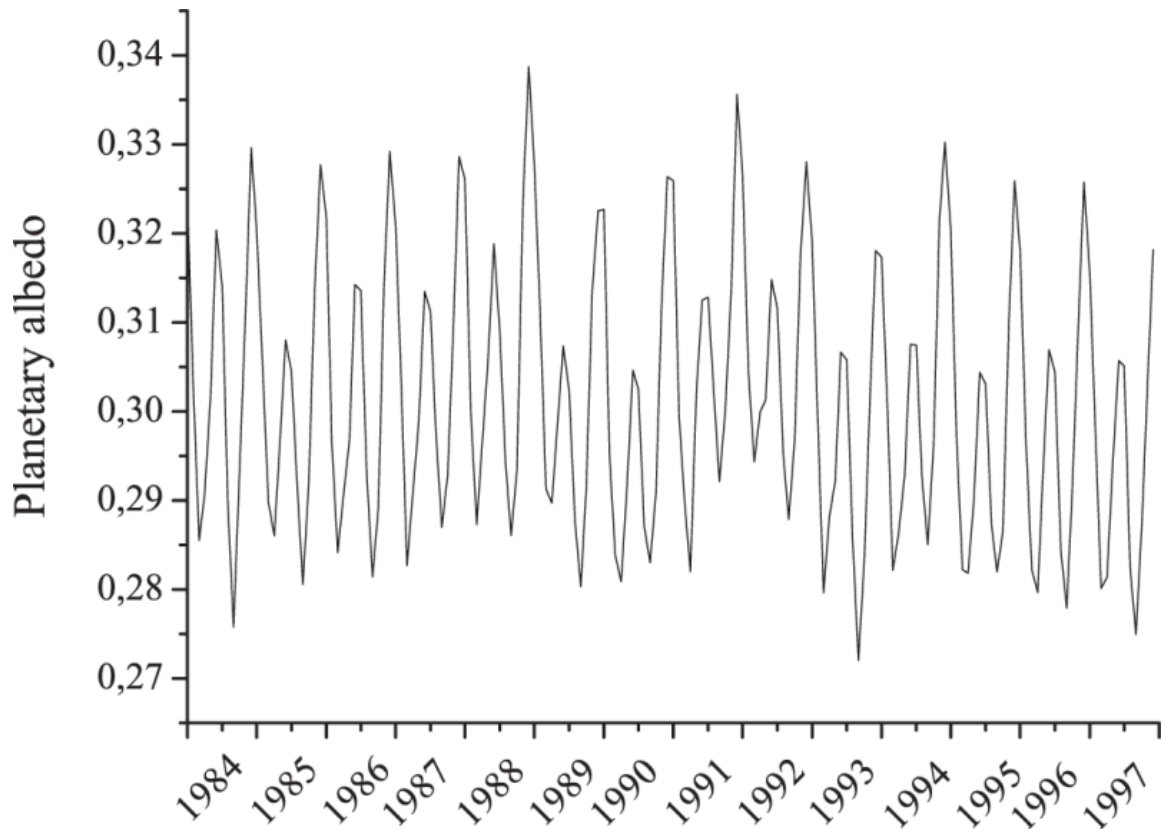

**Figure 7.** Long-term (1984-1997) time series of monthly averaged planetary albedo (adopted from Vardavas and Taylor, 1987).

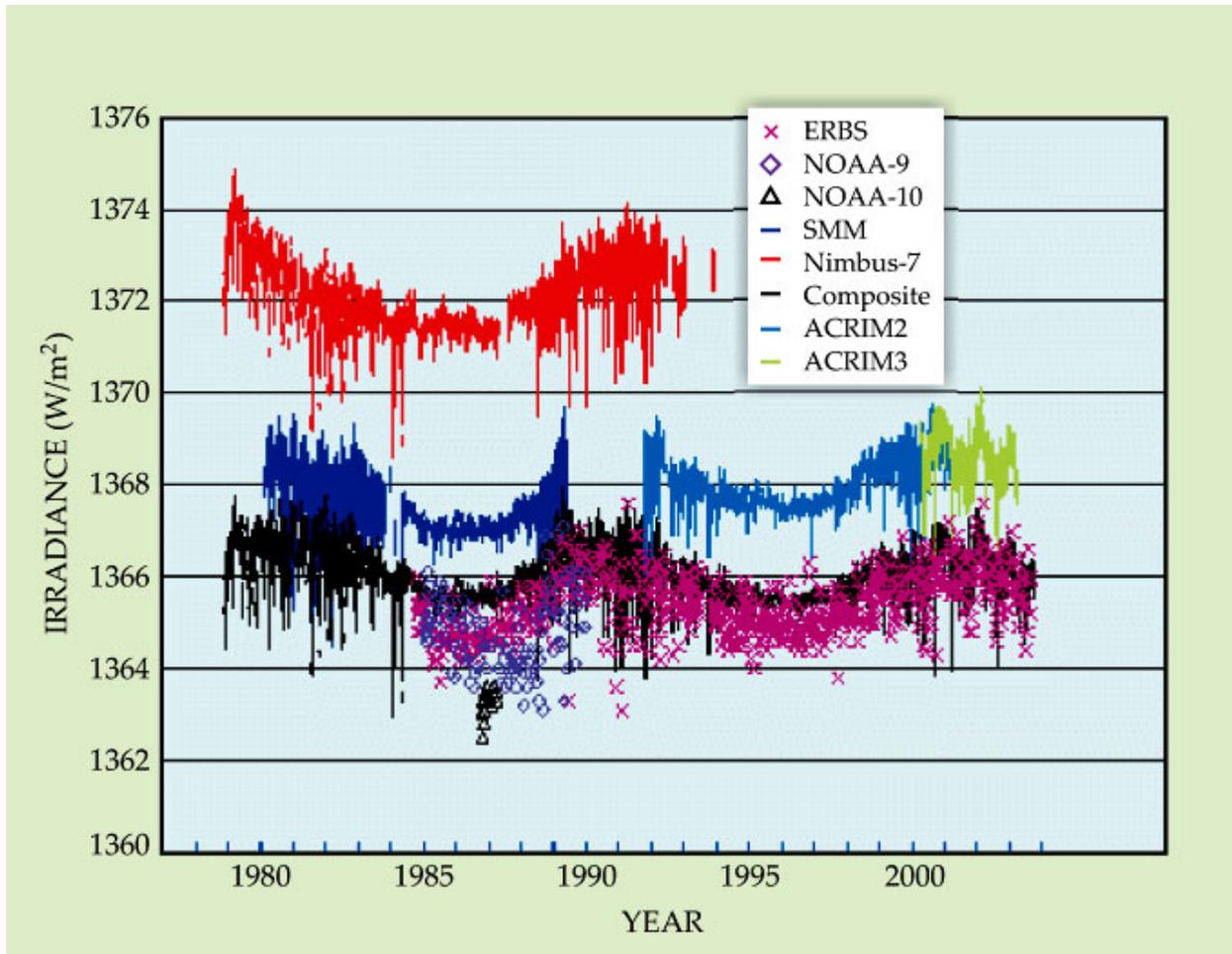

**Figure 8**. As in Figure 5, but without SORCE/TIM (adopted from Duffy et al., 2009). Note that during that time ARCRIM3 was not corrected.